\documentclass[article]{emulateapj}

\makeatletter
\let\@dates\relax
\makeatother

\usepackage{amsmath}

\shorttitle{G2+G2t as an outflow?}

\begin{document}

\renewcommand{\thefootnote}{\alph{footnote}}

\title{The G2+G2t complex as a fast and massive outflow?} 
\author{A. Ballone\altaffilmark{1,2}, M. Schartmann\altaffilmark{1,2,3}, A. Burkert\altaffilmark{1,2,4}, S. Gillessen\altaffilmark{2}, P. M. Plewa\altaffilmark{2}, R. Genzel\altaffilmark{2}, O. Pfuhl\altaffilmark{2}, F. Eisenhauer\altaffilmark{2}, T. Ott\altaffilmark{2}, E. M. George\altaffilmark{2}, M. Habibi\altaffilmark{2}} 

\altaffiltext{1}{University Observatory Munich, Scheinerstra{\ss}e 1, D-81679 M{\"u}nchen, Germany; aballone@mpe.mpg.de}
\altaffiltext{2}{Max-Planck-Institute for Extraterrestrial Physics, Postfach 1312, Giessenbachstra{\ss}e, D-85741 Garching, Germany}
\altaffiltext{3}{Centre for Astrophysics and Supercomputing, Swinburne University of Technology, Hawthorn, Victoria 3122, Australia}
\altaffiltext{4}{Max-Planck Fellow}

\begin{abstract}
Observations of the gas component of the cloud G2 in the Galactic Center have revealed its connection to a tail (G2t) lying on the same orbit. More recent studies indicate a connection between G2 and G1, another cloud detected on the blueshifted side of G2's orbit, suggesting a scenario in which G2 is a denser clump in a stream of gas. In this Letter we show that a simulation of an outflow by a central source (possibly a T Tauri star) moving on G2's orbit and interacting with a hot atmosphere surrounding SgrA* can have G2 and G2t as a byproduct. G2 would be the bow-shock formed in the head of the source, while G2t might be the result of the stripping of the rest of the shocked material by the ram pressure of the surrounding hot gas and of its successive accumulation in the trailing region. Mock position-velocity diagrams for the Br$\gamma$ emission for this simulation can indeed reproduce the correct position and velocity of G2t, as well as the more tenuous material in between. Though some tension between the observations and the simulated model remains, we argue that this might be due to issues in the construction of observed position-velocity (PV) diagrams and/or to a poor treatment of some physical processes - like hydrodynamic mixing - in our simulation. 

\end{abstract}

\vspace{1.4cm}

\section{Introduction}

The small cloud G2, discovered by \citet{Gillessen12} at few thousands Schwarzschild radii ($R_s$) from the supermassive black hole (SMBH) in the Galactic Center, has triggered a large debate on its nature, origin and future evolution. Many of these questions are still unsolved. G2 is detected in L' and M' MIR bands and in few recombination lines like Brackett-$\gamma$ (Br$\gamma$), Paschen-$\alpha$ and Helium I. It has been found to move on a very eccentric orbit, $e\approx 0.98$, and its pericenter passage occurred in early 2014 \citep{Gillessen13b, Phifer13, Pfuhl15}.   
In a series of observations from 2004 to present days, \citet{Gillessen12,Gillessen13a,Gillessen13b} and \citet{Pfuhl15} have shown the tidal stretching of the gas component of G2 unfolding year by year \citep[see however][]{Valencia-S15}. The last observations also provide indications of G2 being just a denser clump in a longer filament on an eccentric orbit around SgrA*. Position-velocity (PV) diagrams show that G2 is followed by a tail (G2t) and linked to it by a ``bridge'' of more tenuous gas \citep[perhaps emitting also in L', as reported by][]{Pfuhl15}. A more careful inspection of the PV diagrams, L' and Br$\gamma$ maps from 2004 to 2008, led to the realization that another clump, G1, is preceding G2 by roughly 13 years on a very similar orbit with lower angular momentum. This finding suggested a link between the three objects, with G1 being part of the same stream of gas containing G2 and G2t and having lost a large fraction of its angular momentum during its pericenter passage \citep{Pfuhl15,McCourt16}. The idea of G2 being part of a larger complex had been previously proposed by \citet{Guillochon14}, where G2 would be a condensation formed in a long stream produced by the tidal disruption of a late-type giant star. This tidal event would have happened during a previous close encounter of such a star with the SMBH in the Galactic Center. A similar stream of clumps might, in principle, also be produced by the collision of winds from binary systems like IRS 16SW \citep{Calderon16}.

On the other hand, as reported blueby \citet{Witzel14}, the bulk of the L' and M' emission from G2 still looks compact close to pericenter, leading these authors to suggest that this dust component at $\approx 560\; \mathrm{K}$ is actually optically thick and internally heated by some stellar object. In particular, G2's dusty component might be the byproduct of a binary stellar merger. However, it is still not clear how such a model might explain G2's gas component. Previous models were able to link the gas component to the dust, in particular to dusty young stellar objects (YSOs): G2 might in fact be a photoevaporating protoplanetary disk \citep{Murray-Clay12} or the wind from a T-Tauri star \citep{Scoville13,Ballone13, DeColle14}.

We present here a 3D adaptive mesh refinement (AMR) simulation of a relatively massive and fast outflow moving on G2's orbit, which reasonably reproduces the G2+G2t complex. Though this model does not take into account G1's detection, it is able to reproduce the main properties of the observed emission from the gas in the redshifted part of G2's orbit. Additionally, this model is simultaneously in agreement with the extended gaseous and compact dusty structure of G2.

\begin{figure*}
\begin{center}
\includegraphics[scale=0.61]{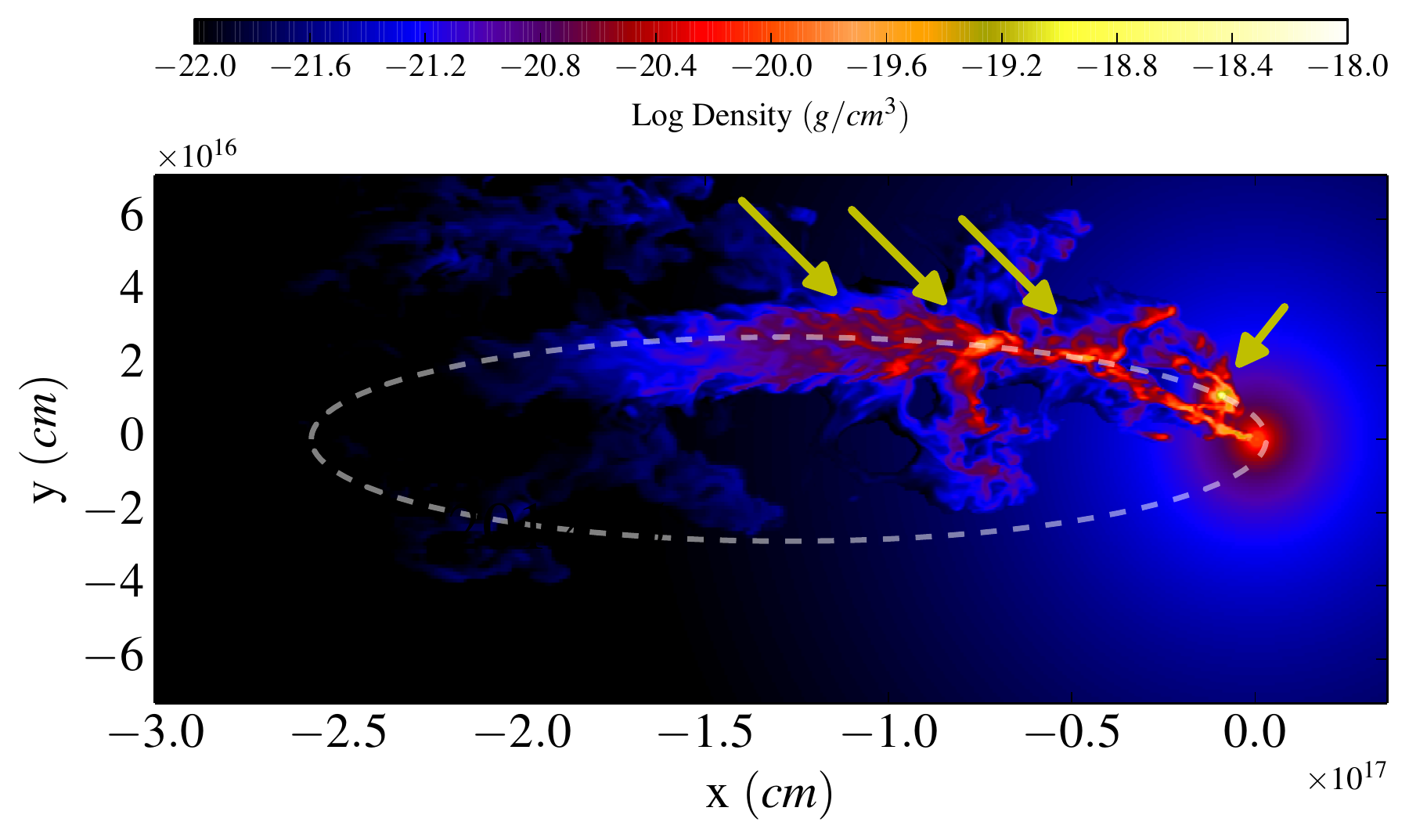}
\includegraphics[scale=0.25]{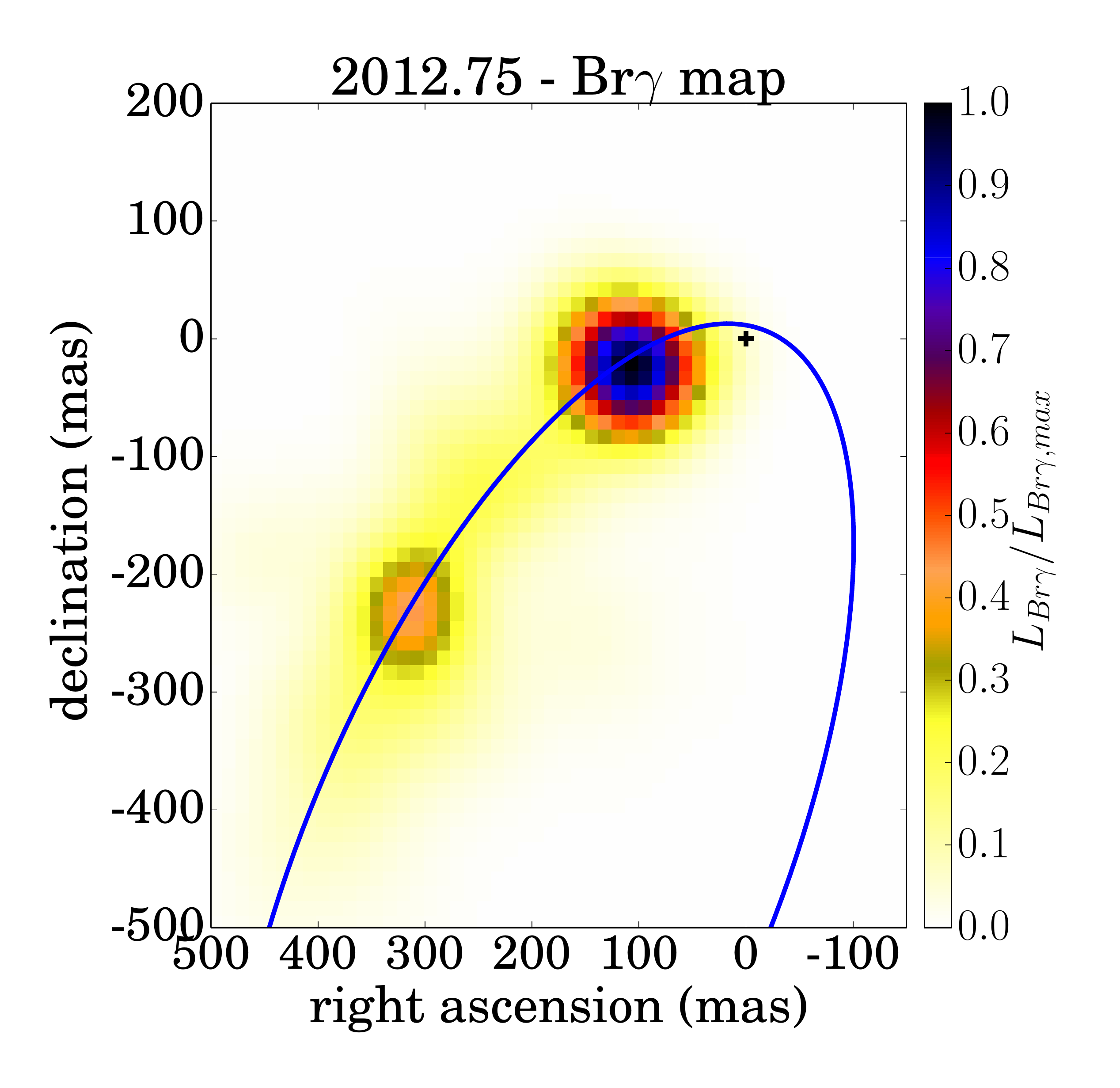}
\caption{Left: Density map for the hydrosimulation in year 2012.75, within the orbital plane $z=0$. The transparent white dashed line marks G2's orbit. The short arrow points at the position of the bow shock, while the three long arrows show the position of the tail. Right: Br$\gamma$ map of the simulation in year 2012.75, projected on the plane of the sky. The blue solid line marks G2's orbit, the black plus sign shows the location of SgrA*. The interaction with the surrounding atmosphere leads to the formation of a two-component structure. 
}\label{simDens}
\end{center}
\end{figure*}

\begin{figure*}
\begin{center}
\includegraphics[scale=0.454]{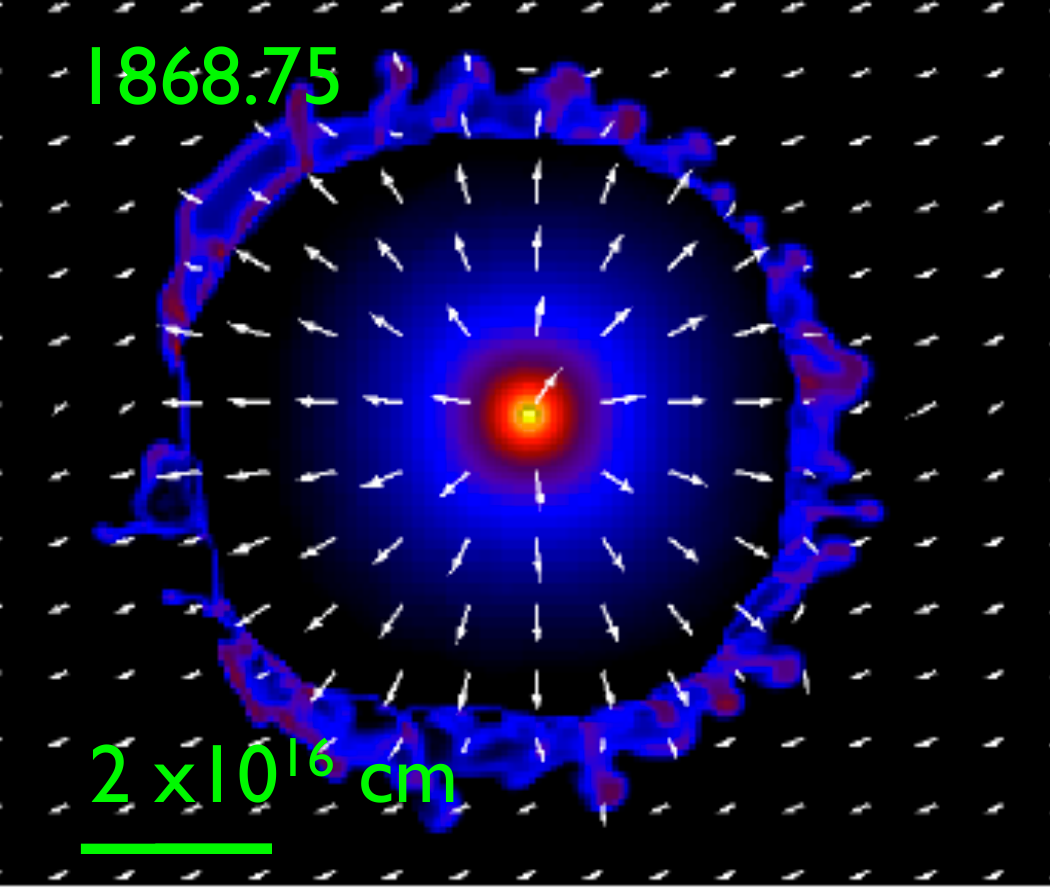}
\includegraphics[scale=0.47]{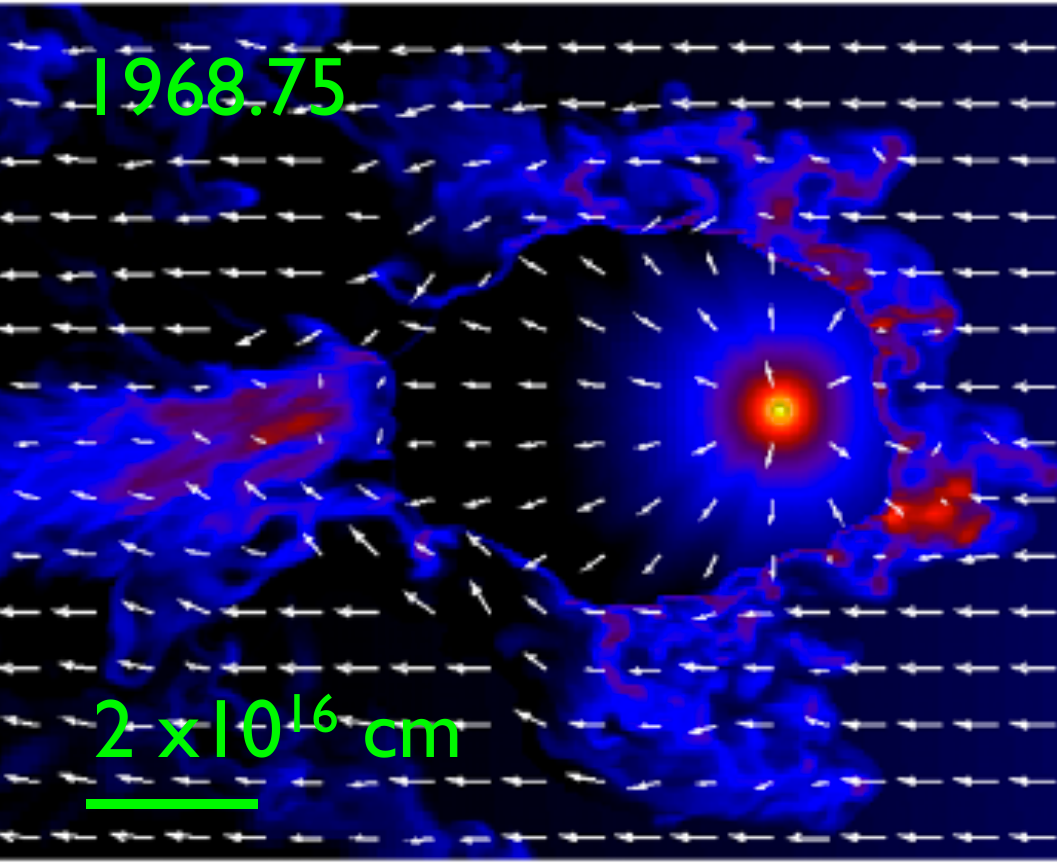}\hspace{0.2cm}
\includegraphics[scale=0.15]{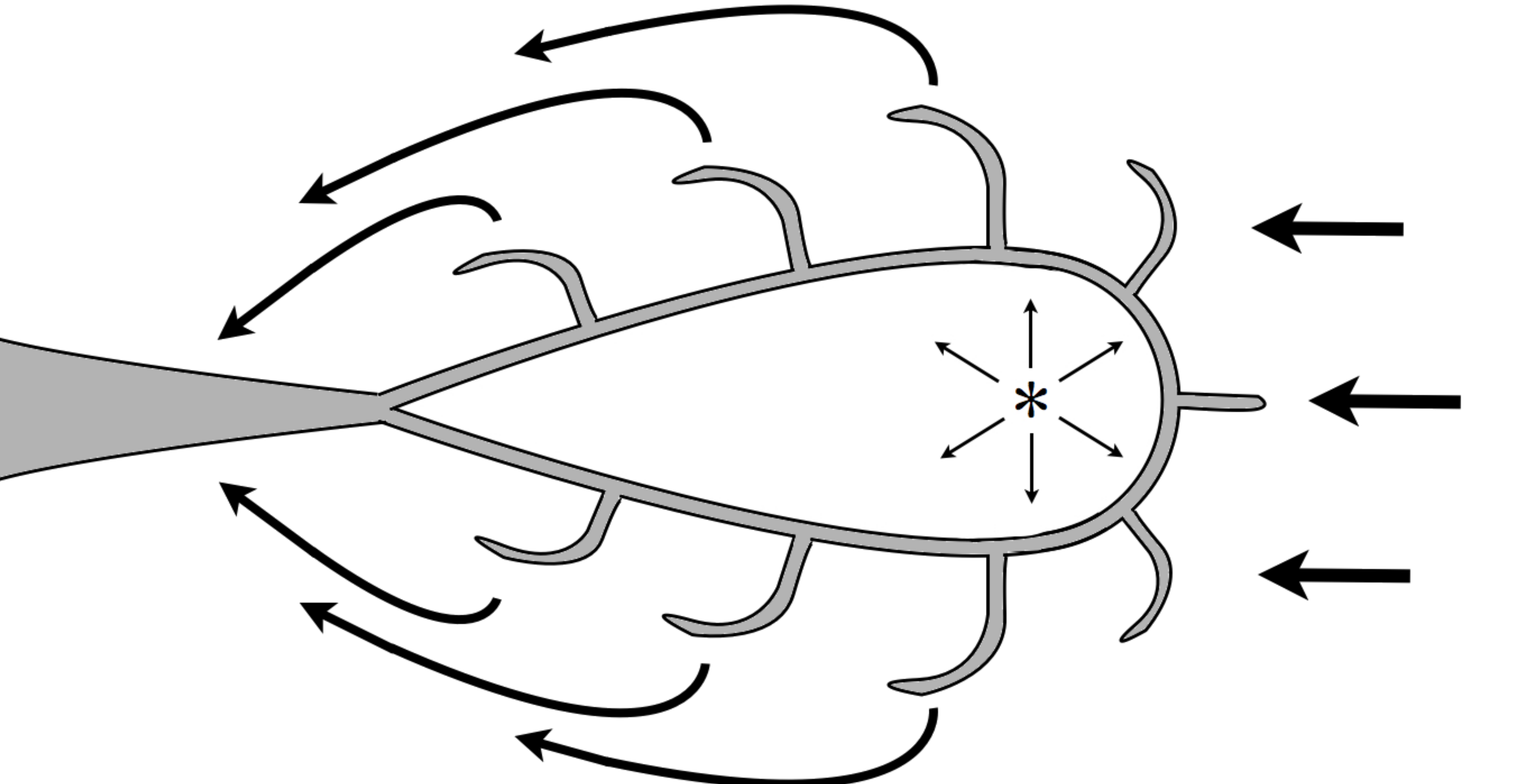}

\caption{Left and central panel: density cuts of our simulation, showing its evolution. The arrows (with arbitrary length) trace the velocity field, after a subtraction of the velocity of the source. The color bar is the same as in Fig. \ref{simDens}. Right panel: sketch of the outflow's evolution at late time (see text). 
}\label{sketch}
\end{center}
\end{figure*}

\section{G2 and G2t as the byproducts of an outflow}\label{secsim}

\subsection{Physical and numerical setup}\label{secsetup}

The simulation in this paper was run with the Eulerian code PLUTO \citep{Mignone07,Mignone12}. Its numerical setup is essentially the same as in \citet{Ballone13}, but here we have moved to a 3D calculation on a Cartesian grid, making use of the AMR strategy implemented in the code. The main update is the implementation of a proper elliptical orbit, derived by \citet{Gillessen13b} through Br$\gamma$ observations. This led to a more realistic evolution of the outflow, and allowed for a more quantitative comparison with the observed PV diagrams (see Sec. \ref{secpv}). The computational domain ranges from $(-3.0\times10^{17};-7.2\times10^{16};-7.2\times10^{16}) \; \mathrm{cm}$ to $(3.6\times10^{16};7.2\times10^{16};7.2\times10^{16}) \; \mathrm{cm}$ in (x,y,z) coordinates, with a finest resolution of $\Delta_{x, y, z, min}=5\times10^{14} \;\mathrm{cm}$.
The outflow was modeled in a ``mechanical'' way; i.e., the velocity and the density were set in a spherical input region following the orbit. The velocity was set to the constant wind value, $v_w$, while the density, $\rho_w$, was set in order to satisfy  

\begin{equation}\label{conteq}
\dot{M}_w=4\pi r_w^2\rho_wv_w,
\end{equation}

where $r_w$ is the distance from the source and $\dot{M}_w$ is the outflow mass-loss rate, kept constant throughout the evolution of the source.

The temperature of the injected material was set to $T_w=10^4 \; \mathrm{K}$ and an adiabatic index $\Gamma=1$ has been assumed, as the cloud is thought to be kept at this temperature by the ionizing photons emitted by the young massive stars in the region and by very efficient cooling \citep[see discussion in][]{Ballone13}.
 
The surrounding hot atmosphere was modeled following the density and temperature distribution of the analytical ADAF model in \citet{Yuan03} and used in several G2 studies \citep[see][]{Burkert12,Schartmann12,Anninos12,Ballone13,DeColle14, Schartmann15}:

\begin{equation}\label{denatm}
n\mathrm{_{at}}\simeq5.60\times10^3 \left(\frac{1}{d_\mathrm{{BH,peri}}}\right)\;\mathrm{cm^{-3}},
\end{equation}
\begin{equation}\label{tematm}
T\mathrm{_{at}}\simeq7.12\times 10^8\left(\frac{1}{d_\mathrm{{BH,peri}}}\right)\mathrm{\; K},
\end{equation}
where $d_\mathrm{{BH,peri}}$ is the distance from SgrA* in units of the distance of G2 at pericenter ($2400 \;R_s$).
As in \citet{Schartmann12, Schartmann15} and \citet{Ballone13}, the atmosphere was reset at each timestep with the help of a passive tracer field, to avoid the onset of convection. 

Finally, the SMBH's gravitational field has been modeled as a Newtonian point source with mass $M_{BH} = 4.31 \times 10^6 M_{\odot}$ \citep{Gillessen09} at $x,y,z = 0$. 

\begin{figure*}
\begin{center}
\includegraphics[scale=0.23]{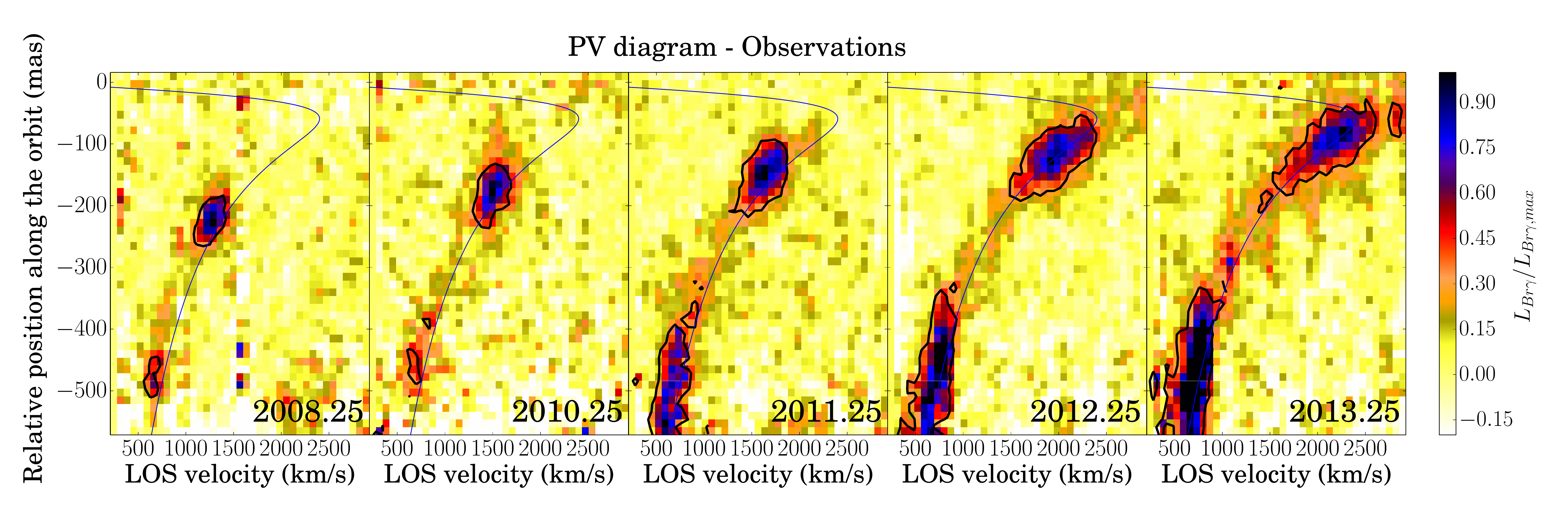}
\includegraphics[scale=0.23]{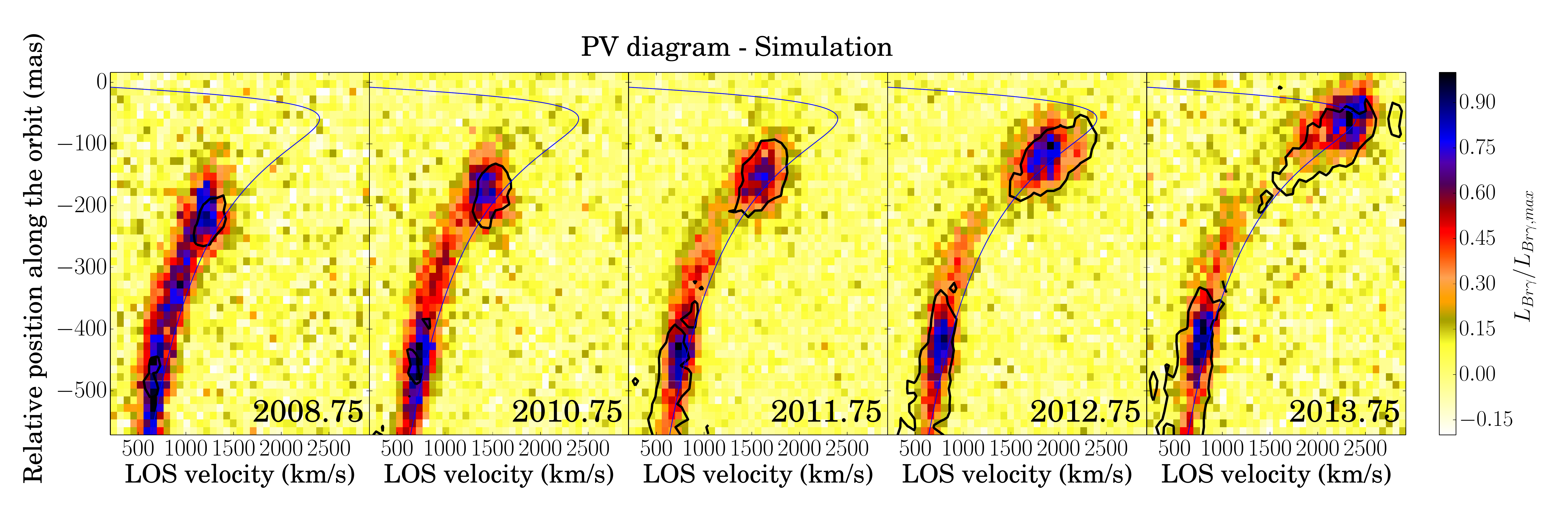}
\caption{Series of observed (top) and simulated (bottom) PV diagrams from 2008 to 2013. The blue line represents G2's orbit as derived from observations, while the black (arbitrary) contours mark the observed positions of G2 and G2t. A time-shift of half a year is applied for the simulations (see text).
}\label{figpv}
\end{center}
\end{figure*}

The outflow and the simulation were started at G2's apocenter. For our orbital solution, this corresponds to year 1818.75. This choice is somewhat arbitrary. However, if the source of G2 had been scattered via several close encounters \citep{Murray-Clay12} in the clockwise rotating disk of young stars \citep{Paumard06, Bartko09}, any pre-existing outflow would not be coherently following the source, being tidally stripped by the encounters.

\subsection{The model}\label{model}

The model has mass-loss rate $\dot{M}_w=5\times10^{-7} \; M_{\odot} \mathrm{yr^{-1}}$ and velocity $v_w=400\; \mathrm{km \;s^{-1}}$. These values are roughly compatible with those of T Tauri stars (see Sec. \ref{secnat}). The left panel of Fig. \ref{simDens} shows a density map in the orbital plane in the year 2012.75. The outflow is composed of a free-flowing region with constant velocity and density declining as $1/r^2$ (cf. Eq. \ref{conteq}). A shock separates this region from the shocked outflow material, which is in a very dense and narrow shell, highly prone to Rayleigh-Taylor instability (RTI). RTI leads to the formation of long fingers; those perpendicular to the orbital motion are easily stripped by the atmosphere's ram pressure. The stripped material, along with the shocked material that is naturally placed there, tends to accumulate in a following tail. This is a general result, in case of efficient ram pressure stripping \citep[e.g.,][]{Pittard05, Vieser07, Cooper09, BandaBarragan16, Christie16} and it is visible in the density cuts and the sketch in Fig. \ref{sketch}. Only when the RTI and the successive stripping are efficient, the outflow results in a clear bimodal density distribution and Br$\gamma$ map (Fig. \ref{simDens}): G2 is produced by the leading termination shock and G2t by the trailing one. 
Whereas the leading standoff distance is given by the ram pressure balance, the position of the trailing termination shock depends on the interplay between several physical processes.

\begin{itemize}
\item \textbf{The thermal pressure of the external medium:} deviating from the classical theory \citep{Wilkin96}, the expansion of an (undisturbed) stellar wind in the backward direction into a high pressure environment stalls when the equilibrium of the wind’s ram
pressure and the atmosphere's thermal pressure is reached \citep{Ballone13, Christie16}.
\item \textbf{The density of the shocked material:} for fixed $\dot{M}_wv_w$, faster outflows are less dens (see Eq. \ref{conteq}) and the same applies to the shocked material. Hence, $\rho_w/\rho_{at}$ is lower, the hydrodynamical instabilities grow faster and the deceleration of the shocked material by ram pressure is higher. The latter has a big role in the amount of material accumulating in the tail, while the former is effecting the position of the trailing termination shock \citep[see][]{Christie16}.
\item \textbf{The strong gravity of the SMBH:} the tidal field is also an important ingredient in affecting the structure of the termination shock, this being defined by momentum balance; furthermore, the tidal force confines the stripped material towards the orbit.
\item \textbf{The time-position dependence of the source velocity and of the atmosphere parameters:} the source moves on a very eccentric orbit and thereby encounters different densities and pressures of the external medium. As a result, the interaction outflow-atmosphere never occurs in a steady-state. E.g., close to apocenter the stripping and the tidal force are always less efficient. If the outflow starts too close to pericenter, the stripping will last for a shorter time and create a less elongated tail. Additionally, the termination shock might still be in an expansion phase (rather than readjust to smaller sizes, as expected for an increase of thermal and ram pressure). If the source has already completed one orbit and it is towards its second pericenter passage, the stripping might have formed a significantly longer tail right before and after the previous pericenter passage. However, the building up of the tail would occur during the whole orbital time ($\simeq 400$ yr) and, in such a timescale, the interaction with the external medium is expected to remove the tail's angular momentum and move it on a significantly different orbit.
\end{itemize}

This complexity clearly expresses the need for hydrodynamical simulations. We refer the reader to the parameter study in \citet{Ballone13}, to understand how the termination shock can be effected by the combined processes discussed before, leading to different separations between the leading termination shock and the tail.

Further ingredients are missing in the present simulation. For example, magnetic fields can suppress the growth of instabilities and the mixing of the tail with the environment. Thermal conduction could instead lead to increased mixing. Resolution has also an effect in these terms. Finally, we must stress that the parameters producing our best model are dependent on the assumed atmosphere, but the latter is also very uncertain. Hence, some caution must be used in considering the constraining power of the present simulation. Still, this represents a needed first step in the investigation of such a scenario.

\subsection{Comparison with the observations}\label{secpv}

The 3D simulation presented here allows us to construct realistic PV diagrams \citep[see][for a detailed description of the used method]{Schartmann15}. In the case of an outflow scenario, the luminosity of the free-flowing region should diverge with decreasing distance $r_w$ from the central source \citep[cf. Eq. \ref{conteq} in this paper and Eq. 5 in][]{Ballone13}. 
However, in the high resolution 2D simulations of \citet{Ballone13} most of the luminosity comes from the shocked material, even when a small inner emitting radius for the free-flowing region is chosen. Additionally, simple estimates show that the shocked shell might be very efficient in ``shielding'' the free-flowing region from ionizing photons coming from the massive stars around G2\footnote[2]{These estimates will be presented in A. Ballone et al. (2016, in preparation).}. Hence, for this analysis, we just consider the Br$\gamma$ emission from the shocked outflow material.

A series of observed and simulated PV diagrams is shown in Fig. \ref{figpv}. Given the complexity of the outflow's structure, the very idealized nature of the simulation and the limited resolution, we restrict ourselves to a simple qualitative comparison. The major feature of the model is the ability to produce a bimodal emission in the PV diagrams, with the two peaks being located roughly at the correct position compared to observations. We find that we must apply a time-shift of half a year between the observations and the output of the simulation for the best match with the observations. This is due to the fact that part of G2's Br$\gamma$ emission is produced by the bow shock surrounding the source in the front but also laterally, so that the source is not exactly placed in the middle of the emitting spot. Given the estimated orbital time of $\approx 400 \;\mathrm{yr}$, this is a minor correction.

Another interesting feature is the presence of more tenuous material connecting G2 and G2t. This is produced by some of the shocked material on the back part of the free-flowing region, as well as by some of the trailing material, though with a smaller flux. 

As a final remark, G2 appears brighter than G2t in simulated Br$\gamma$ maps (see the right panel in Fig. \ref{simDens}), whereas this is not the case in the PV diagrams. This is partially due to the fact that the part of the orbit in which G2 sits has a very small slope in PV space. Hence, in PV diagrams, G2's luminosity is diluted over several velocity bins. G2t's luminosity is also lowered by mixing with the atmosphere, as discussed in the next section.

In conclusion, the simulation can reproduce quite well the two emitting spots and the separation between them. However, there exist differences between the observations and the simulations in the relative luminosity of G2 and G2t. We find that the simulated tail is brighter than observed at early times, while it is slightly underluminous at late times. These issues will be discussed in the next section.

\section{Discussion and conclusions}\label{secdisc}

\subsection{How to reconcile the observed and the simulated PV diagrams?}

Though the global structure of the G2+G2t complex is nicely reproduced by the simulation, a significant mismatch is present in terms of relative brightness of G2t with respect to G2 (see Fig. \ref{figpv}). According to the last observationally derived PV diagrams, G2t \textit{appears} to be flaring up. In fact, a simple dynamical argument is sufficient to rule out the absence of G2t, in 2008 and 2010, from the region of the PV space at positions $<350 \;\mathrm{mas}$: G2t can not dynamically ``enter'' the PV diagrams in the short time interval between 2010 and 2011 observations. However, one of the main reasons for this mismatch might reside in the observational techniques used to extract PV diagrams. The position along the orbit is determined using a curved slit, currently following G2's best-fitting orbit \citep[see][for details on the method]{Gillessen13a,Schartmann15}. The obtained brightness of the tail depends on the chosen slit. For example, G2t's flaring could be explained by slightly different orbits of G2 and G2t \citep[for an indication of this, see Fig. 3 in][]{Pfuhl15} and G2t might have been progressively ``entering'' the slit, particularly at late times. More extended observations and analysis are needed to determine the exact structure of the tail and its connection to G2 (P. M. Plewa et al., 2016, in preparation).
For our model, however, all the emitting material is projected onto the orbit by construction, so it will always fully appear in the simulated PV diagrams.

An explanation for the lower simulated emission of G2t at late times probably resides in the mixing between the outflow material and the atmosphere. 
The mixing of outflow material with the hot atmosphere decreases its density and increases its temperature, thus reducing its overall luminosity \citep[see][for a more detailed discussion on this issue]{Schartmann15}. This effect is more severe for G2t, which is the result of the accumulation of stripped shocked material. However, mixing depends critically on various uncertain parameters, as discussed in Sec. \ref{model}.

\begin{figure}[!h]
\begin{center}
\includegraphics[scale=0.24]{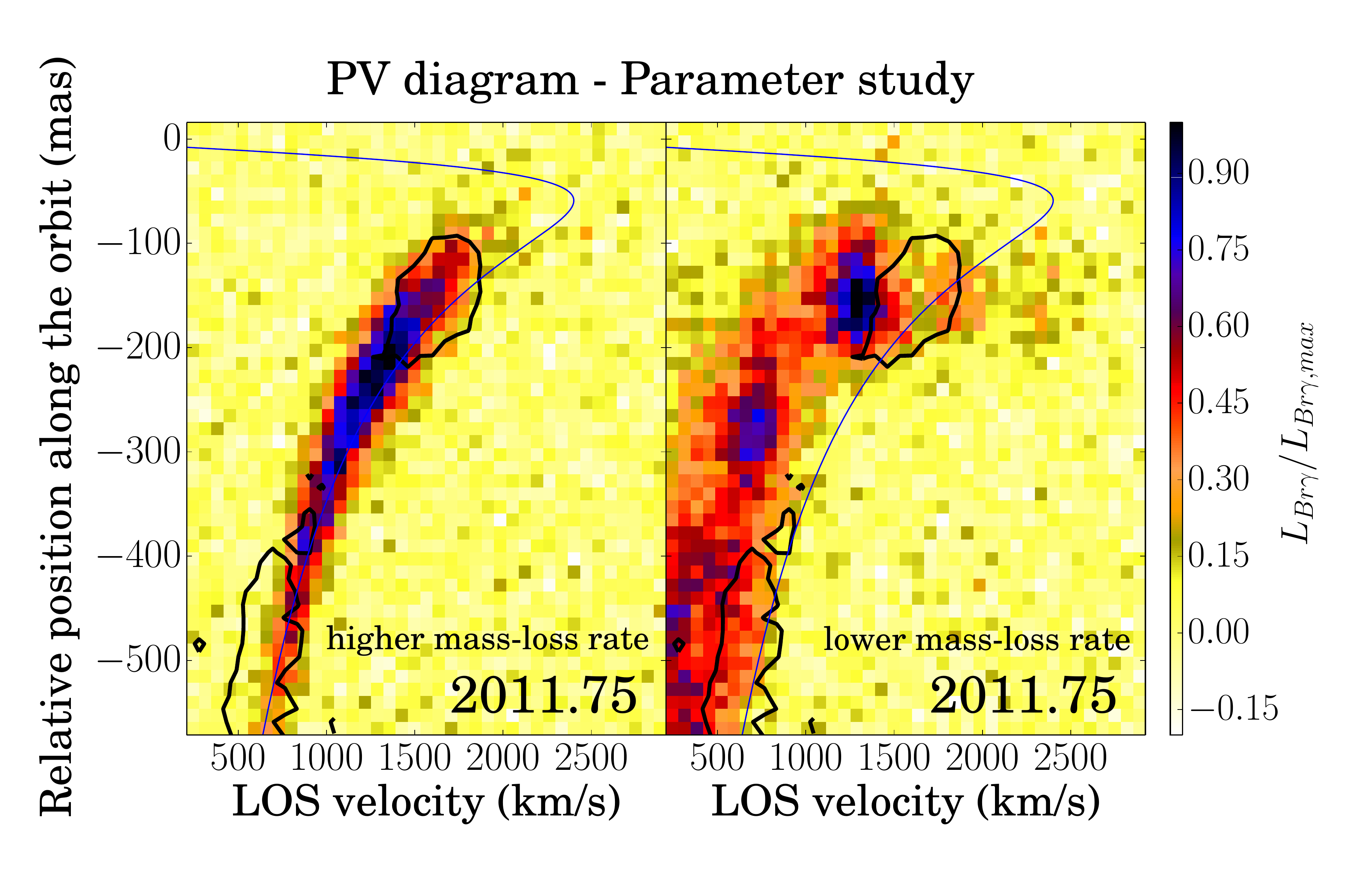}
\caption{Simulated PV diagrams for the year 2011.75 for model A (left panel) and model B (right panel) of our parameter study. Neither is able to account for the observations.
}\label{paramstudy}
\end{center}
\end{figure}

\subsection{The nature of the source}\label{secnat}

A small parameter study showed that the model presented here might be the best (and unique) model able to match the observations for our assumptions. In this parameter study we kept the quantity $\dot{M}_w v_w$ constant and we increased and decreased $\dot{M}_w$, producing two models with $\dot{M}_w=2\times 10^{-6} \;M_{\odot}\mathrm{yr^{-1}}$, $v_w=100\; \mathrm{km\;s^{-1}}$ (model A) and $\dot{M}_w=2\times 10^{-7} \;M_{\odot}\mathrm{yr^{-1}}$, $v_w=1000\; \mathrm{km\;s^{-1}}$ (model B), respectively. PV diagrams for the year 2011.75 are shown in Fig. \ref{paramstudy}. In the case of model A, the free-flowing region is larger and stripping is not efficient enough to form a tail. As a result, the object appears larger in the PV diagrams, but no bimodal distribution of the emitting material occurs. For model B, the stripping of the shocked material is more significant and the size of the free-flowing region is reduced in the trailing side by the backflow. As a result, the separation between the head and the tail is not as a sharp as in our reference model and all the emitting material is shifted to lower velocities compared to the orbit derived for G2 and G2t. We then infer that only relatively small variations of the parameters, around our standard model, are allowed.

As in \citet{Scoville13} and \citet{Ballone13}, the most reasonable source associated with the obtained parameters ($\dot{M}_w=5\times10^{-7} \; M_{\odot}\mathrm{yr^{-1}}$, $v_w=400\;\mathrm{km\;s^{-1}}$) is a T Tauri star. Observations put the wind parameters for T Tauri objects in the ranges $\dot{M}_w=\mathrm{[10^{-12},10^{-7}]} \;M_{\odot} \mathrm{yr^{-1}}$ and $v_w = \mathrm{[50,300] \;\mathrm{km\;s^{-1}}}$ \citep{White04}, placing our favorite model at the boundary of the distribution or even beyond. However, as discussed, many parameters of our model are uncertain and this can reflect on the outflow parameters. Exceptionally massive and fast outflows do anyway exist \citep[as in the case of DG Tau;][]{Guenther09, White14} and relatively high mass-loss rates are probably needed to reproduce the observed luminosity (and estimated mass) of G2. Hence, the most plausible candidate needed for such a scenario would be in any case a T Tauri star, since higher mass stars or different evolutionary states would result in extremely different outflow parameters or they would be detected by the current instruments.

\subsection{Advantages and disadvantages of this scenario}

The present model is for the first time able to reproduce both G2 and its trailing tail G2t in a detailed comparison with the observed PV diagrams. If the source is a T Tauri star, it could naturally explain the presence of dust embedded in G2. Furthermore, if the material launched in the outflow is also dusty, it would explain the more moderate MIR emissions seemingly associated to G2t \citep{Pfuhl15}. 

On the other hand, the present model neglects the discovery of the cloud G1, that seems to be associated with the former objects. We stress here that the connection between G1 and the G2+G2t complex is still very speculative and only future observations will clarify this point. Additionally, as discussed in Sec. \ref{secnat}, there are some issues in reconciling the parameters obtained to match the observed PV diagrams with the parameters of a physical T Tauri.

While there is still some tension with the observations, the present model is able to qualitatively (and, partially, quantitatively) reproduce G2 and G2t, offering a valuable potential explanation for these objects.

\acknowledgments

This project was supported by the Deutsche Forschungsgemeinschaft priority program 1573 (“Physics of the Interstellar Medium”). Computer resources for this project have been provided by the Leibniz Supercomputing Center under grant: pr86re.

\end{document}